\documentclass[twocolumn]{svjour3}          
\smartqed  
\usepackage[version=3]{mhchem} 
\usepackage{graphicx}
\usepackage{setspace}
\usepackage{amsfonts,amsmath,amssymb}
\usepackage[dvipsnames]{xcolor}
\usepackage{epsfig}
\usepackage{float}
\usepackage[normalem]{ulem}

\usepackage[numbers,sort&compress]{natbib}
\newcommand{\rev}[1]{#1}

\journalname{EPJ-E}
\begin{document}

\title{Active spheres induce Marangoni flows that drive collective dynamics
}

\titlerunning{Activity-induced Marangoni dynamics}

\author{Martin Wittmann         \and
        Mihail N. Popescu \and
        Alvaro Dom{\'i}nguez \and 
        Juliane Simmchen
}


\institute{
Martin Wittmann \texttt{(contributed equally)}
\at 
Technical University Dresden, Zellescher Weg 19, DE-01069 Dresden, Germany\\
\email{martin.wittmann@tu-dresden.de} 
          \and
Mihail N. Popescu \texttt{(contributed equally)}\\
\texttt{ORCID: 0000-0002-1102-7538}
	  \at
Max Planck Institute for Intelligent Systems, Heisenbergstr. 3, 
D-70569 Stuttgart, Germany\\
\email{popescu@is.mpg.de}
            \and
Alvaro Dom{\'i}nguez \texttt{(corresponding author)}\\
\texttt{ORCID: 0000-0002-8529-9667} 
        \at  
F{\'i}sica Te{\'o}rica, Universidad de Sevilla, Apdo.~1065, 41080 Sevilla, 
Spain\\
Instituto Carlos I de F{\'i}sica Te{\'o}rica y Computacional, 18071 Granada, 
Spain\\
\email{dominguez@us.es} 
            \and          
Juliane Simmchen \texttt{(corresponding author)}\\
\texttt{ORCID: 0000-0001-9073-9770}
        \at
Technical University Dresden, Zellescher Weg 19, DE-01069 Dresden, Germany\\
            Tel.: +49 351 463-37433\\
            Fax: +49 351 463-37164\\
            \email{Juliane.Simmchen@tu-dresden.de}           
}

\date{Received: Date / Accepted: Date}

\maketitle

\noindent

\begin{abstract}
For monolayers of chemically active particles at a fluid interface, collective 
dynamics are predicted to arise owing to activity-induced Marangoni flow even 
if the particles are not self-propelled. Here we test this prediction by 
employing a monolayer of spherically symmetric active \ce{TiO2} particles 
located at an oil--water interface with or without addition of a non-ionic 
surfactant. Due to the spherical symmetry, an individual particle does not 
self-propel. However, the gradients produced by the photochemical fuel 
degradation give rise to long-ranged Marangoni flows. For the case in which 
surfactant is added to the system, we indeed observe the emergence of 
collective motion, with dynamics dependent on the particle coverage of the 
monolayer. The experimental observations are discussed within the framework of 
a simple theoretical mean field model.
\end{abstract}

\section{Introduction}
\label{sec:intro}

The challenge of endowing micro- and nano-sized particles with
motility, without applying external forces or torques, has received
significant interest from both perspectives of applied and basic
science (see, e.g., the insightful reviews in
Refs. \cite{Paxton2006,Posner2017,Gompper2015_rev}). As noted in the
classic paper of Purcell \cite{Purcell1977}, the issue of motility of
such micrometer-sized objects in Newtonian liquids of viscosity
similar to that of water is particularly intriguing and
interesting. For such systems the Reynolds number is very low and thus
the hydrodynamics is governed by Stokes equations while the motion of
the object is in the overdamped regime. Accordingly, such objects
cannot rely on inertia to move steadily (see also the review in
Ref.~\cite{LaugaRev}). Hence, achieving steady directional motion of
colloidal particles requires strategies to break the time-reversal
symmetry of the Stokes equations \cite{Purcell1977}.
\begin{figure*}[!th]
\centering
\includegraphics[width=0.95\textwidth]{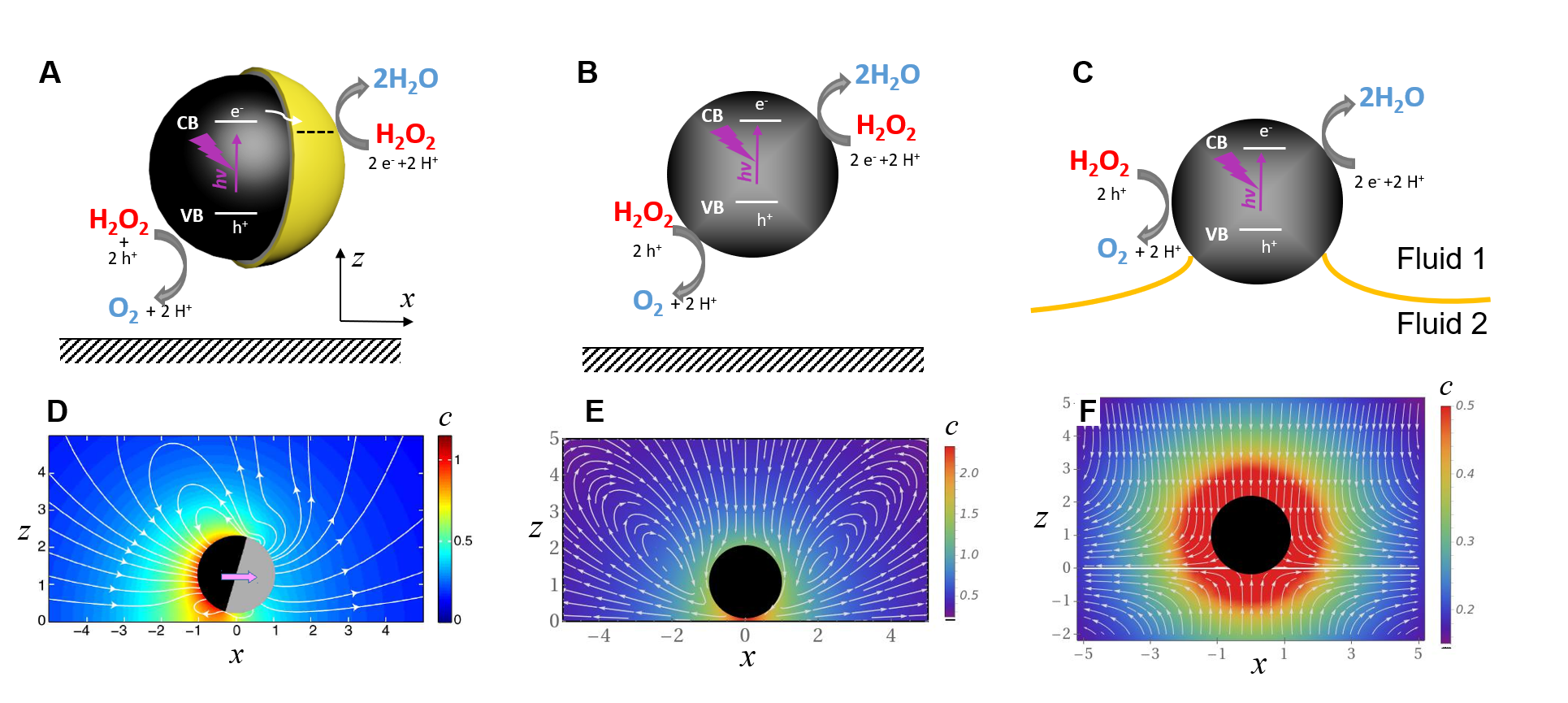}
\caption{(A-C): Schematic representations of a Janus (A) and isotropic 
(B, C) active particles, respectively, in the vicinity of a wall (A, B) or 
at a fluid interface (C). The chemical activity is illustrated via the example 
of a semiconductor photocatalyist promoting, upon suitable illumination, 
the decomposition of hydrogen peroxide (\ce{H2O2}) into water and oxygen. (D-F) 
Examples of the calculated concentration $c$ of reaction product (color coded, 
in arbitrary units) and hydrodynamic flow (streamlines, white).} 
\label{FigIntro}     
\end{figure*}

One direction which has been intensely pursued since the first reports of such 
particles \cite{Paxton2004,Fournier-Bidoz2005}, is making Janus structures. 
These have one part of the surface active (chemically or thermally), while the 
other is inert (see the left panels in Figure \ref{FigIntro}). Variations in 
activity along the surface generate spatial inhomogeneities in the chemical 
composition (or the temperature) of the surrounding suspension.  Motility then 
emerges via, e.g., self-phoresis driven by the gradients of these 
self-generated inhomogeneities 
\cite{Anderson1989,Julicher,Golestanian2005,Kapral2007, Seifert2012a,Koplik2013, 
Michelin2014,Dominguez2020}. The motion of Janus particles, either in unbounded 
fluid or in the vicinity of solid-liquid interfaces has been subject of 
numerous theoretical and experimental studies. Detailed and insightful reviews 
of such studies can be found, e.g., in 
Refs. \cite{Posner2017,Popescu2018,Bechinger_RMP2016}. The question of the 
self-phoresis of active Janus particles near, or trapped at, interfaces between 
two fluid phases, from now on referred to as \emph{fluid interfaces}, has been 
recently tackled both experimentally 
\cite{Stocco,Isa2017,Wurger2017,Isa2018,Sanchez2019,Zohreh2020} and 
theoretically \cite{Sondak_2016,Malgaretti2016,Wurger2017,Malgaretti2018}.

An interesting aspect specific to a fluid interface is that the interface 
itself can respond to the activity of the particles due to the locally induced 
changes in surface tension, which give rise to Marangoni stresses. For a Janus 
particle trapped at the fluid interface, these can directly drive the motion of 
the particle along the interface \cite{Lauga2012,Wurger2014,Masoud2014,IsaMarangoni}. 
Furthermore, the Marangoni flows can couple back to the motion of a 
self-phoretic active Janus particle located \textit{near} the interface 
\cite{Pototsky2019}. Concerning collective effects, for ``carpets'' of 
self-propelled particles at a fluid interface the interplay between motility 
and induced Marangoni flows has been shown to possibly induce instabilities and 
the break up of thin films \cite{Stark2014}.

As can be inferred from the discussion above, most of the studies of  active 
particles have focused on particles for which self-motility is an  intrinsic 
property. The case of chemically or thermally active particles which lack 
motility when isolated, but can set in motion when placed near confining 
surfaces or fluid interfaces, has been somewhat less explored. It was predicted 
that near a fluid interface either self-phoresis \cite{Malgaretti2018} or a 
self-induced Marangoni flow \cite{LGN97,Dominguez2016_a} (as shown schematically 
in the right column of Fig.~\ref{FigIntro}) can move the particle towards or 
away from the planar interface.

When one of these particles with uniform activity over its surface is trapped 
at or near a planar interface, there cannot exist any ``in-plane'' motion, 
because of the radial symmetry of the in-plane Marangoni flow induced by the 
particle (see the right column of Fig.~\ref{FigIntro}) (unless there is a 
mechanism for an in-plane symmetry breaking along the interface, such as, e.g., 
the spontaneous autophoresis at large P{\'e}clet numbers discussed by Ref.~\cite{Michelin2013}). However, when a collection of such particles forms a 
monolayer, the radial symmetry of the in-plane hydrodynamic flow is broken by 
the presence of the other particles\footnote{A similar observation is predicted 
for the case of monolayers of isotropic active particle at a solid-liquid 
interface. However, in this case both the hydrodynamic interactions and the 
phoretic effective interaction (phoretic response of one particle to the 
chemical field produced by another particle) decay at least as $r^{-2}$ with 
the in-plane distance $r$ between particles due to the boundary conditions 
imposed by the solid wall \cite{Popescu2019}. In contrast, the in-plane 
hydrodynamic interactions due to induced Marangoni flows have a significantly 
stronger effect due to their slow decay as $r^{-1}$ \cite{Dominguez2016_a}, 
which is formally analogous to gravity or electrostatics in two-dimensions. It 
can have an attractive or repulsive character, depending on how the surface 
tension reacts to the activity of the particles.}. The superposition of the 
in-plane components of the Marangoni flows induced by each particle becomes an 
effective pair interaction within the monolayer 
\cite{Dominguez2016_a,Dominguez2016_b}\rev{; this is actually a hydrodynamic 
interaction inasmuch as it is due to fluid flow, albeit sourced by the 
interfacial response, rather than by the motion of the particles}. Consequently, 
inhomogeneities in the 
distribution of active particles within the monolayer induce motion of 
the --- otherwise immotile --- particles. Therefore, activity--induced 
collective dynamics within the monolayer may occur in spite of the absence of 
single--particle self-motility. As discussed in 
\cite{Shelley2014,Dominguez2016_b}, at least for simple models it can be shown 
that the state with uniform distribution of particles is unstable against 
perturbations, and collective dynamics sets in driven by the response of the 
interface in the form of Marangoni stresses.\footnote{When the system is also 
laterally confined, as in the case of a sessile drop containing active 
particles, the emergent collective dynamics driven by the self-induced 
Marangoni flows, can be very rich; for example, Ref.~\cite{Fischer2020} 
reported recently the observation of spontaneous symmetry breaking from a 
motionless monolayer to states of self-organized motion of particles and polar 
flow within the drop.}
\begin{figure*}[!th]
\centering
\includegraphics[width=0.9\textwidth]{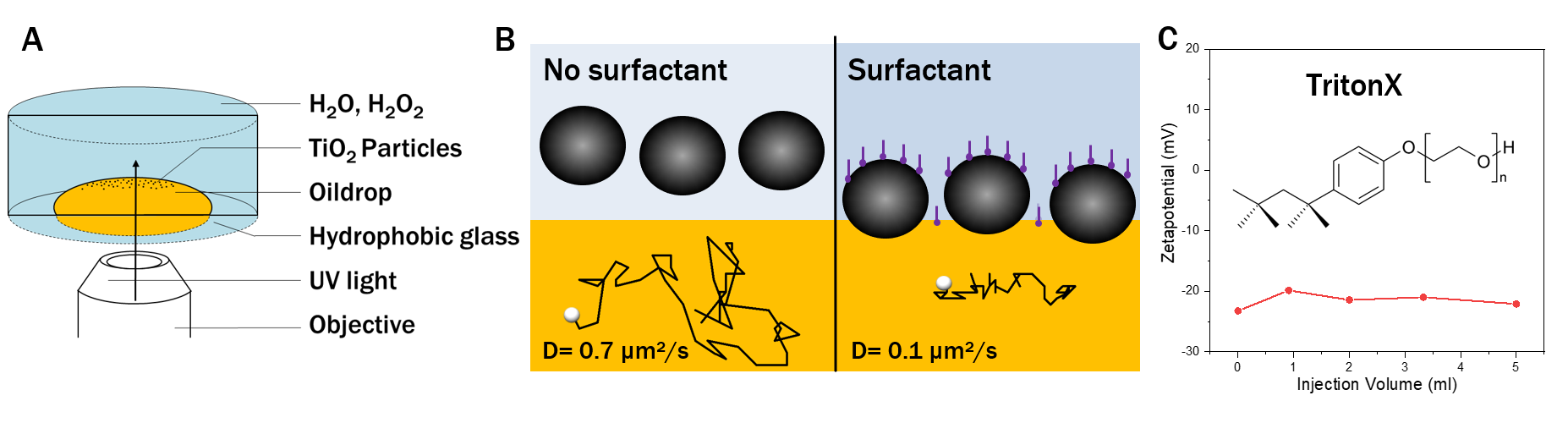}
\caption {A) Schematic representation of the experimental setup. B) Schematic 
representation of the expected location of the particles (based on the 
measured in-plane diffusion coefficient) relative to the 
oil(yellow)--water(blue) fluid interface, without or with surfactant added, 
respectively. The broken lines show typical tracked in-plane trajectories of an 
inactive particle; the corresponding values of the in-plane diffusion 
coefficient $D$ are shown. C) The zeta-potential of particles as a function of 
added surfactant TritonX. The adsorption of the neutral surfactant molecules 
at the particle's surface does not alter the charge conditions, which is in 
contrast to the case of using anionic or cationic surfactants (see 
Fig.~\ref{Zeta} in Appendix~\ref{RefEx}).
}
\label{Setup}   
\end{figure*}

In general, the dynamics within such active monolayers involve the competition 
between the Marangoni effective interaction and other interaction, e.g., the 
direct forces that the particles exert on each other. For example, the issue of 
stability of a monolayer under the competition between the self--induced 
Marangoni flows and the weight--induced capillary attraction, both of which are 
long-ranged (they decay with the distance $r$ between the particles as $r^{-1}$,
for $r$ smaller than the capillary length) was addressed in 
Ref.~\cite{Dominguez2016_b}. The existence of stationary states sustained by 
the competition between the effective Marangoni interaction and a short-ranged 
repulsion was considered in Ref.~\cite{Dominguez2018}.  It has been shown that 
in this case stationary states can emerge, with a radial ``onion-like'' 
structure (from a very dense, solid-like center to a very low, gas-like rim) 
in the particle distribution induced by the presence of the within the 
monolayer, and hydrodynamic flow within the fluid. The estimates for the length 
scale of such spatial structures \cite{Dominguez2018}, in agreement with the 
recently reported experimental set-ups involving monolayers of active colloids
near (or at) fluid interfaces \cite{Isa2017, Isa2018, Wurger2017, Sharan}, 
suggest that experimental validation of those theoretical predictions may be 
possible in the case of a water-air interface and volatile products of the 
chemical activity\footnote{Albeit under the drawback of extremely weak (nm/s) 
Marangoni flows, thus requiring very long experiments, which is technically  
challenging.}. When the Marangoni flows are strong ($ > \mu$m/s), which can be 
the case at an oil--water interface, the mean-field analysis 
in Ref.~\cite{Dominguez2018} cannot be \textit{a priori} justified, and 
interparticle correlations may play a significant role. This hints to the 
exciting possibility of a much richer phenomenology of collective dynamics 
emerging in such systems.

Motivated by these observations, we designed and carried-out a study of the 
emergence of collective dynamics in a monolayer of chemically active spherical 
particles sedimented at a oil--water interface in the presence of a non-ionic 
surfactant. Subsequently, the results are discussed and interpreted in the 
context of previously proposed simple theoretical models of activity-induced 
Marangoni flows 
\cite{Dominguez2016_b,Dominguez2018,LGN97,Shelley2014,Dominguez2016_a}. 

\section{Model system and experimental set-up}
\label{Sec:setup}

As model active particles we use photocatalytically active colloids
based on spherical, isotropic \ce{TiO2} with an average size of 700
nm \cite{doi:10.1021/acsami.9b06128}. Their chemical activity consists in 
promoting the photocatalytic degradation of hydrogen peroxide, \ce{H2O2}, upon 
UV illumination (see the schematic in Fig.~\ref{FigIntro}C). The use of 
photocatalysis has the significant advantage that the chemical activity can be 
switched on and off despite the fuel (\ce{H2O2}) being present in the solution. 
In previous studies \cite{Wang2018} we have shown that very efficient 
self-motile Janus colloids (see Fig.~\ref{FigIntro}A) can be made by 
half-covering such \ce{TiO2} particles with a thin layer of a metal (e.g., 
\ce{Cu}), that alters the reaction rate   locally. Without this additional 
metal layer, only enhanced Brownian motion is observed (see 
Appendix~\ref{RefEx}). When the intensity of UV illumination is sufficiently 
high, a fraction  of the particles lift of from their sedimentation location 
either over a solid wall or over a fluid oil--water interface. 

In order to test the emergence of activity--induced Marangoni dynamics with 
this kind of active particles, we made a specific experimental setup (see 
Fig.~\ref{Setup}A). It consists of a cylindrical cell filled with an aqueous 
solution of \ce{H2O2} (1 \%). This reservoir offers a sufficient quantity of 
fuel in order to guarantee a stable peroxide concentration over the duration of 
an experiment (in the order of several minutes). The bottom wall was 
functionalized with hexadecyltrimethoxysilane in order to achieve hydrophobic
properties and to enhance the tendency of silicon oil to wet it. We used silicon 
oil of viscosity 1000 cSt and mass density 970 $\mathrm{kg/m^3}$, which is 
slightly lower than those of water (1000 $\mathrm{kg/m^3}$) and hydrogen 
peroxide (1100 $\mathrm{kg/m^3}$). An oil--water interface is created by 
carefully depositing with a canula a drop of oil on the hydrophobic bottom wall 
of the cell. Despite the density mismatch, the wetting forces prevent the drop 
from lifting-off by buoyancy and, at the same time, ensure a pancake-like shape 
of the drop, which has proven difficult to achieve using microfluidic chambers 
\cite{Sharan}. These combined effects enable a fluid interface that is flat to 
a good approximation and allows imaging over an extended area, which is crucial 
for the optical microscopy observations. When \ce{TiO2} particles are added 
into the cell, they sediment and a monolayer is formed near this quasi-flat 
fluid interface which will be the focus of our analysis. In  absence of the drop 
of oil the monolayer forms above the substrate solid wall. 

Once the cell has been prepared, it is placed under the microscope and the 
illumination with UV light is performed through the objective, from below. The 
illuminated area has an almost square shape, and its dimensions can be varied 
to a certain extent.  For this study, we fixed it to a rectangular area of 
about 1500 $\mathrm{\mu m^2}$. A consequence of this setup is that only the 
particles that lie inside this area become chemically active. The rest of the 
particles are expected to remain inactive and they serve as passive tracers of 
the flows arising in the aqueous phase.

\section{Results and Discussion}
\label{result}

\subsection{\label{general_dep} Observation of collective dynamics}

First, in order to rule out that the UV radiation \textit{per se} might cause 
changes of the fluid interface that would induce flows, we tested the setup 
with passive \ce{SiO2} particles (see Fig.~\ref{SiO2oil} in 
Appendix~\ref{RefEx}). We observed indeed no macroscopic or collective motion, 
apart from the expected Brownian fluctuations. Second, we tested the setup with 
\ce{TiO2} particles in the absence of either fuel (\ce{H2O2}) or UV 
illumination. Only Brownian motion of the particles was observed. Accordingly, 
from these we infer that particle activity is a prerequisite for collective 
motion.

However, when the particles in the cell without surfactant become active upon 
exposition to both fuel and UV light, again no collective motion is detected 
(see Fig.~\ref{nosurfactantoil} in Appendix~\ref{RefEx}). We only observe, as 
already remarked in Sec.~\ref{Sec:setup}, that some particles within the 
irradiated area manage to escape the focus plane by drifting in the vertical 
direction away from the fluid interface. This observation indicates that the 
particles reside close to the fluid interface, without actually being trapped 
by it. Indeed, it is known that pure hydrophilic particles generally do not 
show much adsorption onto a fluid interface \cite{Ravera}.

Our interpretation of the lack of observable collective motion is  that the 
dynamics may be too weak to be detected, with velocities  that fall under the 
experimental resolution (well below  $0.1\,\mu\mathrm{m/s}$). Therefore, in 
order to enhance the Marangoni flows, we added a surfactant to the aqueous phase 
with the goal to fix the particles at the interface, so that a cumulative 
effect arises due to the superposition of the Marangoni flows by each particle. 
(The addition of surfactant serve to overcome the energy barrier that inhibits 
the attachment of particles to the fluid interface, thus allowing the control of 
the  positioning of the particles relative to the interface 
\cite{binks_horozov_2006}.) The surfactant addidion might also result in an 
increase of the responsiveness of the interface by changing the surface tension, 
however, there is no direct way of measuring this effect. 

After testing different types of surfactants, we opted for TritonX, a non-ionic 
surfactant that does not significantly influence the zeta-potential of the 
particles in order to keep the charge conditions constant (see Fig.~\ref{Zeta} 
in Appendix~\ref{RefEx}). In the absence of surfactant and irradiation, the 
diffusion coefficient of the particles sedimented on the fluid interface is 
about $0.7\, \mu\mathrm{m^2/s}$, which is comparable to the value when the 
particles reside on a solid substrate (i.e., in the absence of the drop of oil). 
After addition of TritonX, the diffusion coefficient decreased significantly to 
a value $0.1\, \mu\mathrm{m^2/s}$. This decrease is due to the influence of the 
higher viscosity in the silicon oil phase \cite{doerr_hardt_masoud_stone_2016}. 
Corroborated by the absence of lift-off when the particles are active, this 
indicates that the particles are effectively trapped at the interface.

Therefore, we repeated the experiment with active \ce{TiO2} particles in the 
presence of a low (0.05 wt\%) concentration of TritonX. Now we observe the 
emergence of collective flows of particles in the plane of the interface (see 
Figs.~\ref{LowConc} and \ref{HighConc}). In order to support that this is due 
to a Marangoni flow, we tested that, when the particles reside on a solid 
substrate, the addition of surfactant does not induce any in-plane collective 
motion (see Figs.~\ref{nosurfactantsolid} and \ref{surfactantsolid} in
the Appendix~\ref{RefEx}). \rev{When the fluid-fluid interface is replaced by a fluid-solid interface we do not observe any collective dynamics  (see Appendix B.5), which convincingly rules out the chemical (``phoretic'') interactions as the source of the observed behavior. 
To dismiss the assumption of thermally driven Marangoni flows (induced by a hypothetical heating of the titania particles due to UV absorption), we compared the behavior of UV-illuminated active particles at the oil-(water plus peroxide) interface in presence and absence of surfactant and found collective dynamics only in the first case. 
Moreover, these same observations allow us to also rule out mechanisms based on Marangoni stresses due to some trace amounts of contaminants at the interface, a scenario invoked by Ref.~\cite{IsaMarangoni} in the context of the motion of heated Au@\ce{SiO2} Janus particles at an oil-water interface.
}

In summary, we have identified the minimal ingredients required for the 
emergence of significant activity--induced collective dynamics with spherically 
symmetric particles in our experimental setup: besides the photocatalytic 
particles fuelled by hydrogen peroxide, and a fluid interface in our specific 
conditions also a surfactant is necessary in order to generate observable 
Marangoni flows.

\subsection{\label{areal_dens_dep} Dependence of the collective
  dynamics on the areal density of particles}

The photocatalytic activity of the \ce{TiO2} particles scales with light 
intensity and fuel concentration, both leading to an increase in product 
concentration. We also confirmed that the behavior of the system depends on 
these two factors, resembling strongly to what is presented here. However, it 
turns out that the impact of the areal density of particles in the monolayer
on the emerging dynamics is much more significant. For this reason, we decided 
to keep the light intensity and the fuel concentration fixed and, for the rest 
of this study, to investigate the dynamics at different particle densities.
\begin{figure}[!ht]
\includegraphics[width = 0.95\columnwidth]{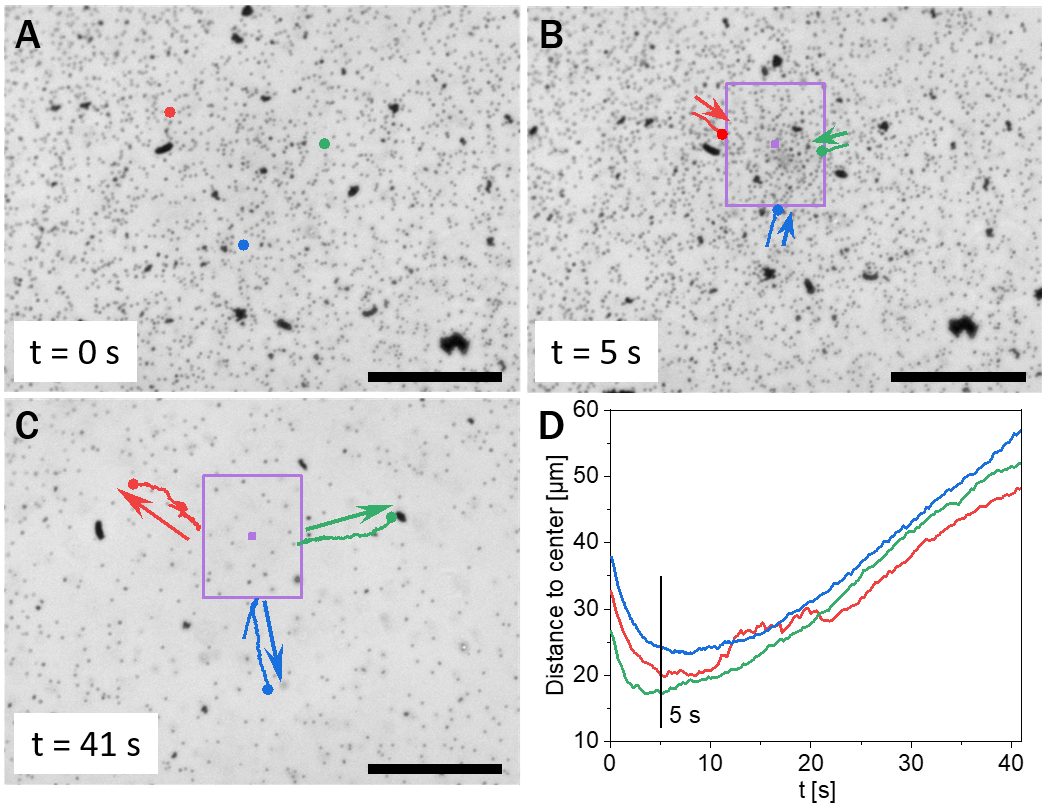}
\caption{Behaviour at low areal fraction: A) Before onset of illumination the 
particles are uniformly distributed over the interface. B) When UV irradiation 
is on (the violet square indicates the illuminated area, but the camera is not 
sensitive to UV light) an inward flow sets in, bringing particles towards (and 
into) the irradiated area. C) The flow reverses direction after approximately 5 
s; the outwards flowing induces a depletion of particles from the illuminated 
area. D) The distance to the center of the illuminated area for some tracked 
particles (the trajectories in B and C) is plotted as a function of the time 
after UV irradiation is switched on. (The scale bars correspond to 50 $\mu$m.)}
\label{LowConc}      
\end{figure}

At low particle areal density (12 \% surface coverage), we observe the effect 
displayed in Fig.~\ref{LowConc}. Before irradiation with UV light, the particles
are uniformly distributed over the fluid interface (Fig.~\ref{LowConc}A). Once 
the UV irradiation in the marked area is switched on, a collective flow towards 
the center of the illuminated region is observed (Fig.~\ref{LowConc}B). After a 
few seconds, and without any changes in the experimental parameters, the 
direction of the flow reverses (Fig.~\ref{LowConc}C). A steady state is not 
reached within the experimental times. When the illumination is turned off, the 
flows stop and the system relaxes to its original, equilibrium state.
\begin{figure*}[!ht]
\begin{center}
\includegraphics[width = 0.9\textwidth]{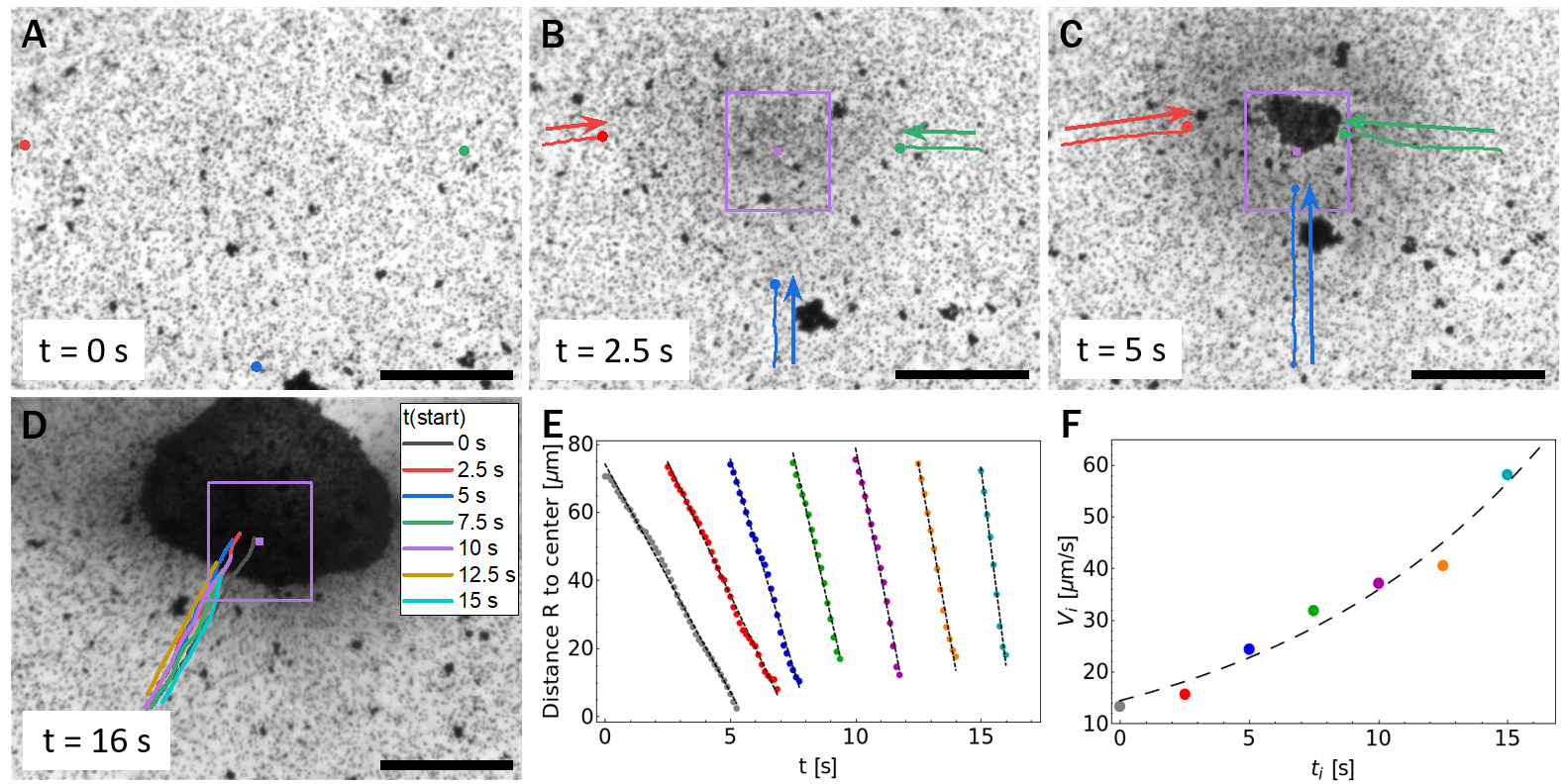}
\end{center}
\caption{Behaviour at high areal fraction: A) Before onset of illumination the 
particles are uniformly distributed over the surface. B) When UV irradiation is 
on (the violet square indicates the illuminated area) an inwards flow sets in 
and brings particles towards (and into) the illuminated area. C) Formation of 
a cluster within the illuminated region. D) As long as the UV light remains 
on, the flow is maintained and the cluster grows. E) The distance to the center 
of the illuminated area for the tracked particles (trajectories shown in panel 
D) as a function of the time after the UV irradiation is switched on. The 
straight lines represent a fit to a motion with constant velocity $V_i$. F) 
Plot of the velocity magnitude $|V_i|$ of a particle which at time $t_i$ (after 
illumination is turned on) is at a given distance $R_i$ to the center of the 
illuminated area (in this example, $R_i\approx 70\,\mu\mathrm{m}$). The dashed 
line is the fit by Eq.~(\ref{eq:Vi}). (The scale bars correspond to 50 $\mu$m.)
}
\label{HighConc}
\end{figure*}

At higher particle concentrations (32 \% surface coverage), the experiment 
starts analogously to the low concentration setup: before irradiation with UV 
light, the particles are uniformly distributed over the fluid interface, albeit
with a higher surface coverage (Fig.~\ref{HighConc}A). Switching on the UV 
irradiation in the area indicated in Fig.~\ref{HighConc} also causes a 
collective flow towards the irradiated area. But, differently to the previous 
case, we do not observe any velocity reversal here. The flow drags the particles
towards the center of the illuminated area, where they form a cluster (see 
Fig.~\ref{HighConc}D). For the trajectories shown in Fig.~\ref{HighConc}D, 
corresponding to particles passing through the same location at different times 
after the UV was turned on, the velocity for each tracked trajectory is 
approximately constant (see Fig.~\ref{HighConc}E). Comparing these velocities, 
it can be inferred that the magnitude increases with the time passed since the 
illumination was switched on (compare the slopes of the lines in  
Fig.~\ref{HighConc}E, from left to right).  Alternatively, by considering a 
fixed time, e.g., $t = 4\,\mathrm{s}$, it can also be inferred that the 
velocity of a particle is larger when its distance from the illuminated area is 
larger. This inflow leads to a steady growth of the cluster; we do not observe 
particles escaping the cluster as long as the UV light is on. After turning off 
the UV light, the flows also stop; however, the dense cluster remains compact 
and merely breaks apart into larger pieces as the system relaxes. This fact can 
be interpreted as indicative of Marangoni flows that are sufficiently strong to 
push the particles so close as to foster van der Waals forces causing an 
essentially irreversible aggregation.

In summary, we can state that the observed collective flows depend 
significantly on the coverage by active particles. We now attempt to frame 
these findings into a simple theoretical model for activity--induced Marangoni 
flows.

\subsection{\label{theo_arg} Theoretical model and analysis}

Based on the experimental analysis, which pinpoints the activity-induced 
Marangoni flows as the most plausible source of the collective dynamics within 
the monolayer, the interpretation of the results is attempted within the 
theoretical framework proposed in Refs.~\cite{Shelley2014,Dominguez2016_a}. 
This is the simplest model for the collective motion of active particles by the 
self-induced Marangoni flow. Succinctly, it treats the monolayer as a continuum
dragged by the ambient flow. This flow is described in the Stokes 
approximation, and it is driven by the chemical gradients determined according 
to the diffusion equation (see Fig.~\ref{FigIntro}F). \rev{The evolution 
within the monolayer is thus due to the hydrodynamic interactions between the 
particles (sourced by the interfacial response); even though the particles are 
lacking self-motility, the hydrodynamic interactions are long ranged ($\sim 
1/r$, see, c.f.,  Eq.~(\ref{eq:fieldu})) because of the induced Marangoni 
flows.}

Thus, by adopting a coarse--grained approach to describe the large scale
dynamics, we employ continuum fields that are assumed to vary slowly over the 
microscopic length scales (size of the particles, mean interparticle 
separation, etc\dots). The monolayer plane is identified with $z=0$; the vector 
$\vec{r}=(x,y)$ denotes the in-plane position and $\nabla := 
\left(\partial_x,\partial_y \right)$ the two--dimensional (2D) nabla operator 
in the monolayer plane. The areal number density of particles in the monolayer 
is given by the field $\varrho(\vec{r},t)$, which is the only relevant field 
because the particles are not intrinsically motile (i.e., there are no fields
associated to a polar or a nematic order). We assume that there is no particle 
flux in or out of the monolayer; accordingly, $\varrho(\vec{r},t)$ is a 
conserved quantity and obeys the continuity equation
\begin{equation}
  \label{eq:cont}
  \frac{\partial \varrho}{\partial t} = - \nabla\cdot (\varrho \vec{u}) ,
\end{equation}
as the particles are dragged by the ambient Marangoni flow $\vec{u}(\vec{r})$ 
induced by the activity of the particles, i.e., we assume that the only 
relevant cause of particle motion is this flow\footnote{The model can be easily 
extended to account for other driving forces, e.g., the thermal (Brownian) 
forces and the direct interactions between particles (e.g., steric repulsion), 
see Ref.~\cite{Dominguez2018}.}. Notice that, although the Marangoni flow 
exists in the bulk of the fluid phases, the only relevant contribution to the 
dynamics is the 2D flow evaluated at the monolayer plane. A particularly 
important consequence is that, although the three-dimensional (3D) flow is 
incompressible, the projection of the flow onto the monolayer plane \emph{is} 
compressible (see, c.f., Eq.~(\ref{eq:nablau})), and thus Eq.~(\ref{eq:cont}) 
does describe the emergence of an inhomogeneous distribution within the 
monolayer.

The activity of the particles is modeled as a source term in the concentration 
of a chemical involved in the reaction (e.g., \ce{O2} as product, or \ce{H2O2} 
as reactant). One can assume that the molecular diffusion is much faster than 
the time scales associated with the collective motion, so that the distribution 
of chemical is adapted to the instantaneous configuration of the particles. 
Consequently, the concentration field can be found from the 3D Fick's law with 
sources due to the active particles, i.e., by neglecting the time dependence
as well as the drag by the ambient flows (low P{\'e}clet number), and solving a 
Poisson equation --- see the color-coded field in Fig.~\ref{FigIntro}F.

Since the surface tension of the interface depends on the local chemical 
composition, spatial variations of the surface tension develop due to the 
inhomogeneous distribution of chemicals in the fluid media. These gradients in 
surface tension translate to tangential forces at the interface (Marangoni 
stresses), which are transmitted to the fluids and generate the Marangoni flow. 
Assuming again that the flow adapts instantaneously to the particle 
configuration, the associated 3D velocity field is described by the Stokes 
equations for incompressible flow (low Reynolds number) --- see the white 
streamlines in Fig.~\ref{FigIntro}F.

Finally, one assumes that the surface tension changes linearly with the local 
concentration, which is a reasonable hypothesis if the range of variations is 
not too large. When this assumption is combined with the solutions of Fick's 
law for the concentration of chemical and of Stokes' equation for the flow, one 
ends up with a simple relationship between the particle density of the 
monolayer and the Marangoni flow at the monolayer plane (see Ref.~\cite{Dominguez2018} for the detailed derivation):
\begin{equation}
  \label{eq:nablau}
  \vec{u} = - \nabla \Phi ,
  \;\;
  \nabla\cdot\vec{u} =
  \left\{
    \begin{array}[c]{cl}
      G \varrho(\vec{r}),
      & \vec{r}\in \textrm{illuminated area,}
         \\
      & \\
      0 ,
      & \vec{r}\not\in \textrm{illuminated area,}
    \end{array}
  \right.
\end{equation}
where the constant $G$ (which can be either positive or negative) is 
proportional to the activity of the particles\footnote{More precisely, if the 
activity of a particle is quantified by the time rate $Q$ at which the chemical 
reaction proceeds, then $G= Q b/[2\pi(\eta_1+\eta_2)(D_1+\lambda D_2)]$. Here 
$b$ is the proportionality coefficient between the changes in the surface 
tension and the changes in the chemical concentration (in the linear 
approximation), $\eta_1, \eta_2$ are the viscosities of the fluid phases, $D_1, 
D_2$ are the diffusivities of the chemical in the fluids, and $\lambda$ is the 
ratio of solvabilities.}.
  
Equations~(\ref{eq:cont}) and (\ref{eq:nablau}) form a closed system that 
allows one to obtain the monolayer density $\varrho(\vec{r},t)$. Notice that 
Eq.~(\ref{eq:nablau}) means that the 2D flow at the monolayer is actually a 
Newtonian field; e.g., for $G<0$ the dynamics could be read as the (overdamped) 
collapse of a mass distribution under its own 2D gravity. The equations can be
solved exactly, under the assumption of radial symmetry, for the initial 
condition of a homogeneous distribution, $\varrho(\vec{r},0)=\varrho_0$ (see
Appendix~\ref{app:theory}). In particular, for $G<0$ (flow ``falling'' into the 
illuminated area) the model predicts a 2D radial Marangoni flow in the 
non--illuminated area with radial component 
\begin{equation}
  \label{eq:fieldu}
  u_r(r,t) = - \frac{A \mathrm{e}^{t/T}}{2\pi T r} ,
\end{equation}
where $T=(|G|\varrho_0)^{-1}$ is a characteristic time scale set by the
activity, and $A$ is the area of the illuminated region. 

One can compare this result with the quantitative observations presented in
Fig.~\ref{HighConc} as follows. During the observation time, the velocity of a
particle does not change significantly (see Fig.~\ref{HighConc}E), so that it
would be given by its initial value: a particle located at a distance $R_i$ in
the non-illuminated region at the time $t_i$ (measured from the onset of the 
illumination at $t = 0$) will have a velocity
\begin{equation}
  \label{eq:Vi}
  V_i = - C_i \mathrm{e}^{t_i/T} ,
  \qquad
  C_i = \frac{A}{2\pi T R_i}.
\end{equation}
This expression is used to fit the experimental data points, as shown in 
Fig.~\ref{HighConc}F; it can be seen that it provides a good approximation 
with the two fitting parameters $T\approx 11\,\mathrm{s}$, $C_i\approx 
14.4\,\mu\mathrm{m/s}$. With these numbers, Eq.~(\ref{eq:fieldu}) also predicts 
that the velocity of a particle is indeed approximately constant during the 
observation time (see Eq.~(\ref{eq:Verror}) in Appendix~\ref{app:theory}).

In spite of these results, there are discrepancies indicating that the 
experimental system is too rich to be fully captured by the simple theoretical 
model we have employed. First, the combination $T C_i \approx 
158.4\,\mu\mathrm{m}$ of the fitting parameters differs significantly from the 
prediction $A/2\pi R_i \approx 3.4\,\mu\mathrm{m}$ for the value $R_i\approx 
70\,\mu\mathrm{m}$ of the trajectories depicted in Fig.~\ref{HighConc}E and 
the area $A\approx 1500\,\mu\mathrm{m}$ of the illuminated region. Second, and 
more importantly, at given time $t$ the velocity field according to 
Eq.~(\ref{eq:fieldu}) decays with the distance $r$, while the trajectories in 
Fig.~\ref{HighConc}E, which are probing this velocity field, actually show a 
flow that at fixed $t$ increases in magnitude with $r$. This means, in 
particular, that this flow is compressible even in the non-illuminated regions, 
in disagreement with Eq.~(\ref{eq:nablau}). Additionally, the theoretical model 
cannot explain the trajectories with velocity reversal depicted in
Fig.~\ref{LowConc}D, because the sign of the velocity is fixed by the
sign of the constant $G$ (see Eq.~(\ref{eq:ur}) in Appendix~\ref{app:theory}).

These arguments suggest that, contrary to the model assumptions, the activity 
that drives the flow is not located only in the illuminated region. The 
observation that, in the absence of surfactants, no collective motion is 
detected, allows one to further conjecture that the surfactant is sensitive to 
the chemical reaction at the active particles, e.g., by desorbing from the 
\ce{TiO2} particles when they are active or by reacting with produced 
intermediate reactive oxygen species. Accordingly, one may attempt to extend the 
theoretical model by allowing for additional sources of Marangoni flow, so that
$\nabla\cdot\vec{u}\neq 0$ also in the non-illuminated region. In any case, 
these findings raise new questions, whose answer would require further 
experimental studies and complementary theoretical modeling.

\section{Conclusions}

In conclusion, we have set up an experimental study to test the prediction of 
emergent collective dynamics driven by activity induced Marangoni flows, 
rather than self-propulsion, within a monolayer of chemically active particle 
at a fluid interface. The setup involves \ce{TiO2} particles, which under UV 
illumination promote the photocatalytic decomposition of hydrogen peroxide in 
aqueous solutions. The particles are spherical by construction, so that they 
lack self-motility. We studied the configuration of particles sedimented at a 
quasi-flat silicon oil--water interface, in the presence of the neutral 
surfactant TritonX in the aqueous phase. Upon UV illumination of a small 
central region, we have indeed observed the emergence of radial flows dragging 
the particles in the monolayer towards or away from the illuminated region, 
depending on the areal particle density within the monolayer.

Through a set of complementary experiments, we could identify activity induced 
Marangoni flows as the most plausible cause of this dynamics and rule out 
various other a priori possible mechanisms (such as, e.g., phoretic interactions 
or UV-response of the surfactant). Surprisingly, the dynamics exhibits a 
significant, qualitative change upon increasing the average areal density in 
the monolayer. At low densities, an initial transient inflow towards the 
illuminated region is replaced, within few seconds, by a monotonic outwards 
flow, so that a growing area around the illuminated region emerges depleted of 
particles. In contrast, at large areal densities the inflow persists, leading 
to the steady growth of a particle cluster in the illuminated region.

The trajectories tracked for individual particles were analyzed within the 
framework of the simplest mean--field model for the dynamics driven by activity 
induced Marangoni flow \cite{Dominguez2018}. It turns out that the experimental 
setup is very rich and cannot be captured fully by the model. The most relevant 
conclusion is the finding that, contrary to expectations, the interfacial 
stresses that drive the Marangoni flow are not localized in the illuminated 
area, but rather they also exist outside, although the photochemical activity 
in the illuminated area remains the driving factor. This led to the conjecture 
that the role played by the surfactant is more than just facilitating the 
entrapment of the particles at the interface or providing a more responsive 
interface. For instance, one can conceivably argue that the surfactant, which 
is not restricted to stay in the illuminated area, may be sensitive to the 
photochemical reaction at the \ce{TiO2} particles. Future studies are required 
to understand and rationalize these findings.

\begin{acknowledgements}
JS and MNP thank the SPP 1726 for providing a platform for scientific
exchange and the Freigeist Fellowship grant number 91619 
from Volkswagen foundation. AD acknowledges support from the Spanish Government through Grant
FIS2017-87117-P, partially financed by FEDER funds.
\end{acknowledgements}

   


\appendix

\section{Experimental Details}

\subsection{Materials}

\underline{Chemicals:}
All chemicals and solvents were used in analytical grade without any additional 
treatment: titanium (IV) isopropoxide (Alfa Aesar Co. Ltd.); dodecylamine 
(Fluka); silicone oil (Sigma-Aldrich). 

\noindent\underline{Particle synthesis:} TiO$_{2}$ particles were synthesized 
via an improved version of a previously published method \cite{Wang2018}. In 
brief, 0.18 mL of water was added to mixture solution of 105 mL methanol and 45 
mL acetonitrile.  Then 0.28 g of DDA was dissolved in the mixture and stirred 
for about 10 min. 1 mL of TTIP was added dropwisely and stirred for 12 hours, 
precipitates were removed and stirring was continued for 24 hours, when the 
process was repeated until no further precipitates were observed. Then the 
particles were washed with methanol three times and were calcined in a tubular 
furnace under nitrogen flow for 2 hours at 600 $^\circ$C.
 
\noindent\underline{Surface functionalization:} 24x24 mm glass slides were 
washed by sonication in aceton and ethanol and subsequently plasma cleaned. The 
surface functionalization was performed in the gas phase in a desiccator at 5 
mbar for two days using 30 $\mu$L hexadecyltrimethoxysilane per glass slide.

\subsection{Methods}

\begin{figure}[H]
	\centering
	\includegraphics[width=0.9\columnwidth]{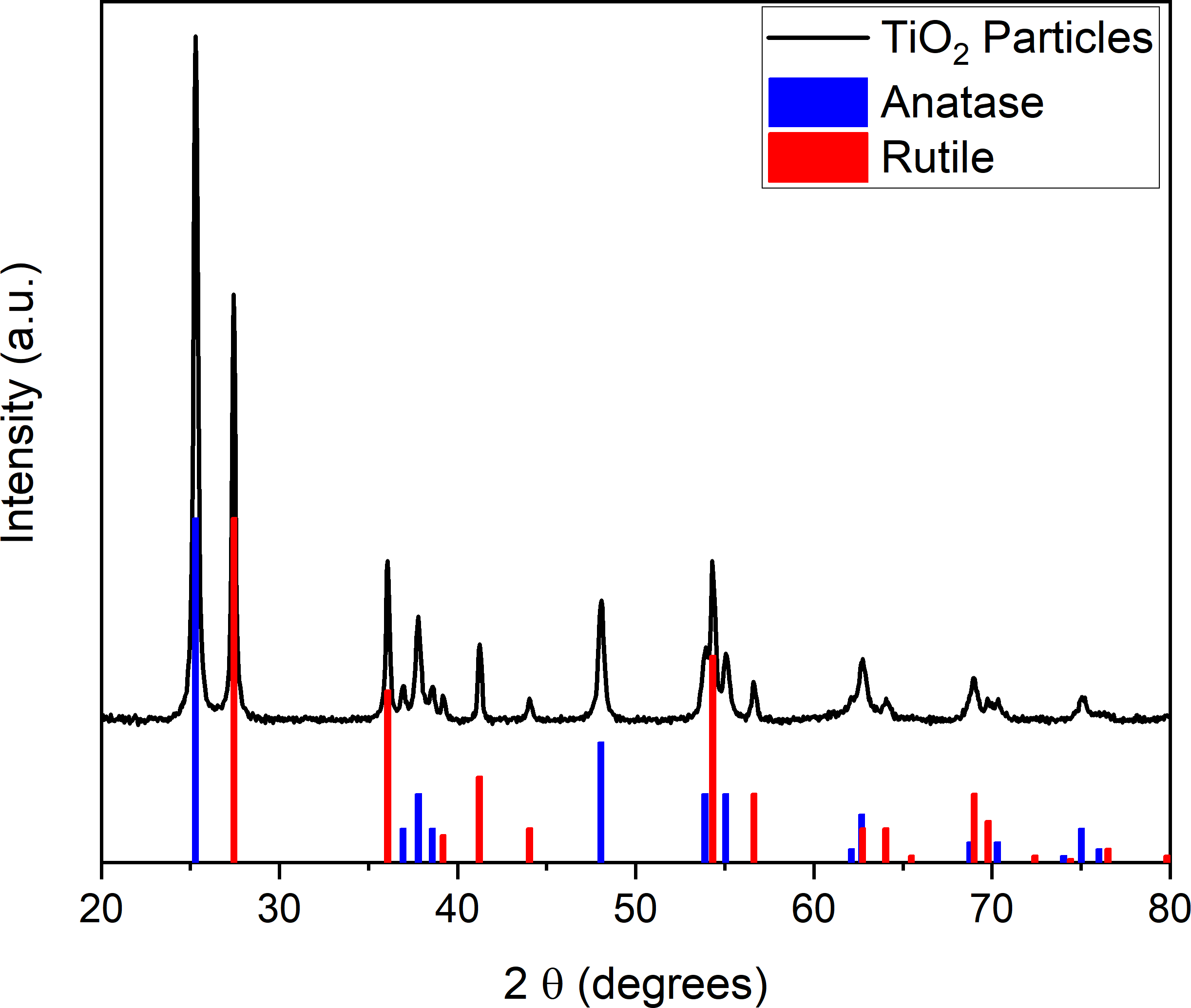}
	\caption{XRD of \ce{TiO2} particles.}
	\label{XRD}
\end{figure}

\noindent\underline{XRD:} The phases of the \ce{TiO2} particles were identified 
using XRD. Measurements were carried out with a Bruker 2D phaser in a 2$\Theta$ 
range of 20$^\circ$--80$^\circ$. 
The XRD of \ce{TiO2} particles (\ref{XRD}) only includes reflexes corresponding 
to the anatase and rutile phases, confirming the presence of a mixture of both 
polymorphic forms of \ce{TiO2}.

\noindent\underline{Zeta-potential measurements:} The zeta-potential was 
measured with a Malvern Zetasizer Nano ZSP in autotitration mode.

\noindent\underline{Video recording:} An inverted optical microscope (Carl 
Zeiss Microscope GmbH) equipped with a Zeiss Colibri LED lamp and a 
``N-Achroplan'' 63x/0.95 M27 objective were used for observation. The 
wavelength of the UV light was 385 nm, the UV lamp power was fixed at 100\% 
lamp intensity, corresponding to 315 mW. The particle behavior was recorded 
with a Zeiss camera (Axiocam 702 Mono) and a frame rate of 40 fps.

\noindent\underline{Experiments on liquid interfaces:} The cylindrical cell was 
filled with 800 $\mu$L of a solution containing 1 \% \ce{H2O2} and 0.05 wt\% 
TritonX. A drop of silicon oil was placed on the hydrophobic bottom using a 
canula.  10 - 50 $\mu$L of particles dispersed in water (10 mg/mL) were added to 
the cell and given time to settle down before the observation with the 
microscope was started.

\noindent\underline{Reference experiments on solid substrate:} The
experiments on solid substrate were carried out in the same cylindrical cell 
without the oil drop.
   
\noindent\underline{Video analysis:} Videos were analyzed using ImageJ 1.52p. 
Tracking was done with the Manual Tracking plugin.

\section{Reference experiments}
\label{RefEx}

\subsection{Reference experiments on a fluid interface using passive \ce{SiO2} 
particles}

\begin{figure}[!bht]
	\centering
	\includegraphics[width=0.99\columnwidth]{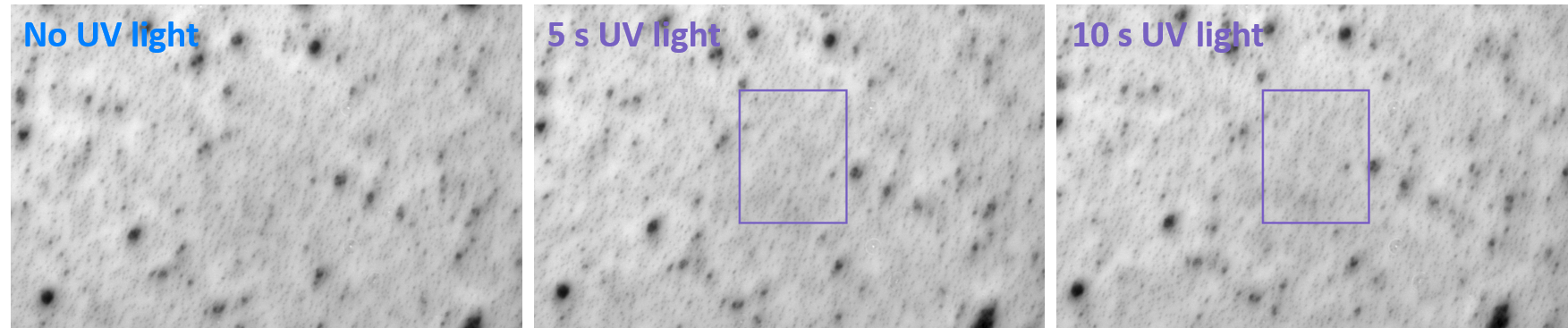}
	\caption{Passive \ce{SiO2} particles in a solution of TritonX
          and \ce{H2O2} on oil--water interface. Left: before UV
          irradiation. Middle: after 5 s of UV irradiation. Right:
          after 10 s of UV irradiation.}
	\label{SiO2oil}
\end{figure}

In the experiments with passive silica particles on the solid substrate and on 
the fluid interface, no motion induced by the UV light was observed (see 
Fig.~\ref{SiO2oil}).

\subsection{Reference experiments on a fluid interface without surfactant 
addition}

\begin{figure}[!bht]
	\centering
	\includegraphics [width=0.99\columnwidth]{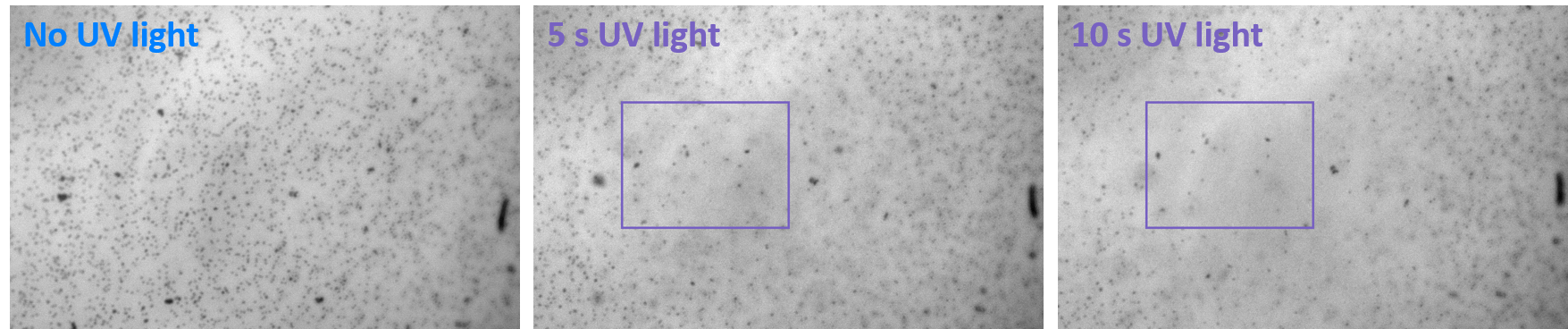}
	\caption{\ce{TiO2} particles without addition of surfactant on
          oil--water interface. Left: before UV irradiation. Middle:
          after 5 s of UV irradiation. Right: after 10 s of UV
          irradiation.}
	\label{nosurfactantoil}
\end{figure}

In these experiments we have observed that irradiation with UV light causes 
enhanced Brownian diffusion of the \ce{TiO2} particles, but does not lead to 
collective dynamics. Some particles lift off and move upwards out of focus (see 
Fig.~\ref{nosurfactantoil}) if the UV intensity is sufficiently high.

\subsection{Reference experiments on a fluid interface without \ce{H2O2} addition}

\begin{figure}[!thb]
	\centering
	\includegraphics[width=0.99\columnwidth]{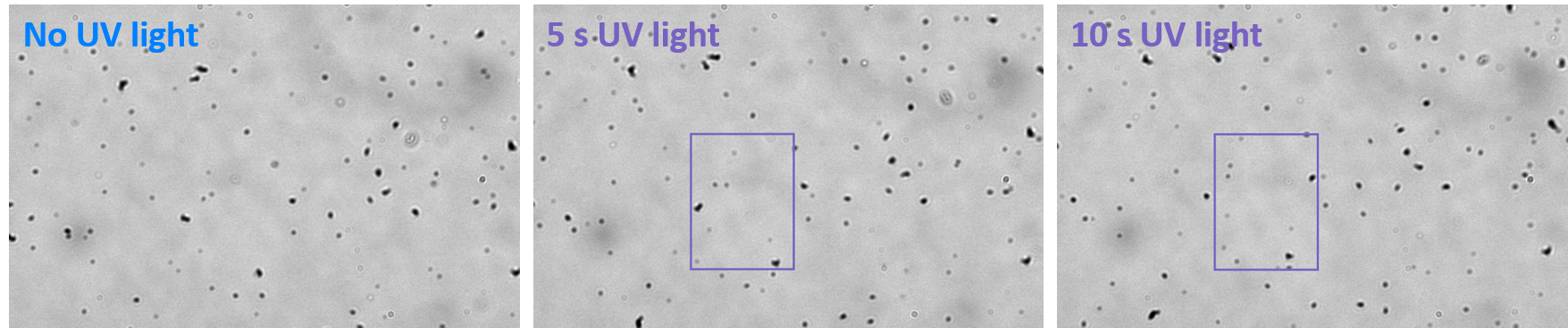}
	\caption{\ce{TiO2} particles on oil--water interface in a
          solution of TritonX without \ce{H2O2}.  Left: before UV
          irradiation. Middle: after 5 s of UV irradiation. Right:
          after 10 s of UV irradiation.}
	\label{noperoxideoil}
\end{figure}

If \ce{H2O2} is not added to the solution, the UV light does not induce any 
particle motion (see Fig.~\ref{noperoxideoil}).

\subsection{Choosing a surfactant --- Zeta-potential measurements}

Zeta-potential titrations were performed using Malvern Zetasizer Nano ZSP in 
autotitration mode using the attached Multi Purpose Titrator. The titration was 
carried out adding the respective surfactant (SDS, CTAB and TritonX) and 
zeta-potential values were recorded after every ml of the solution of 
surfactant was added.

\begin{figure}[H]
	\centering
	\includegraphics[width=0.99\columnwidth]{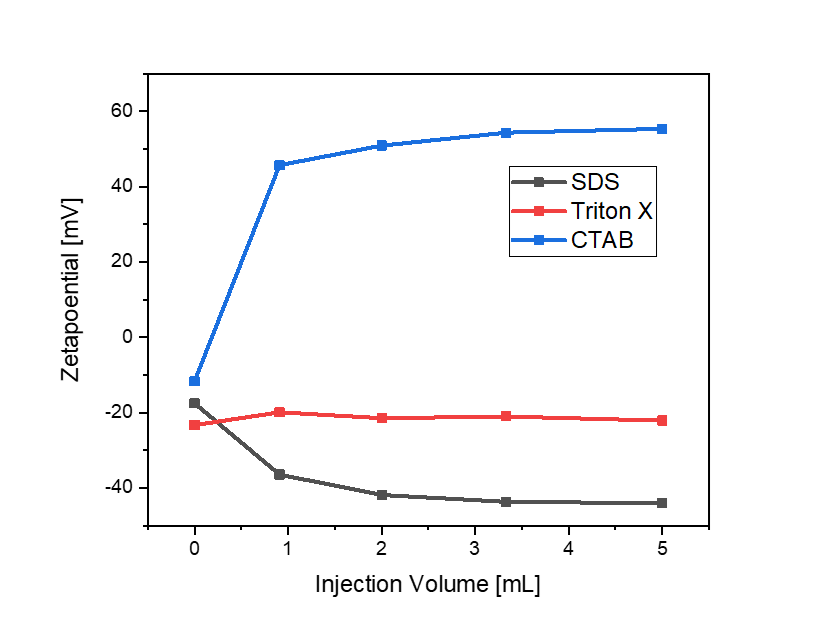}
	\caption{Zeta-potential titrations of \ce{TiO2} particles with
          different surfactants (SDS, CTAB and TritonX).}
	\label{Zeta}
\end{figure}

\subsection{Reference experiments on solid substrates}

\begin{figure}[!thb]
\centering
\includegraphics[width=0.99\columnwidth]{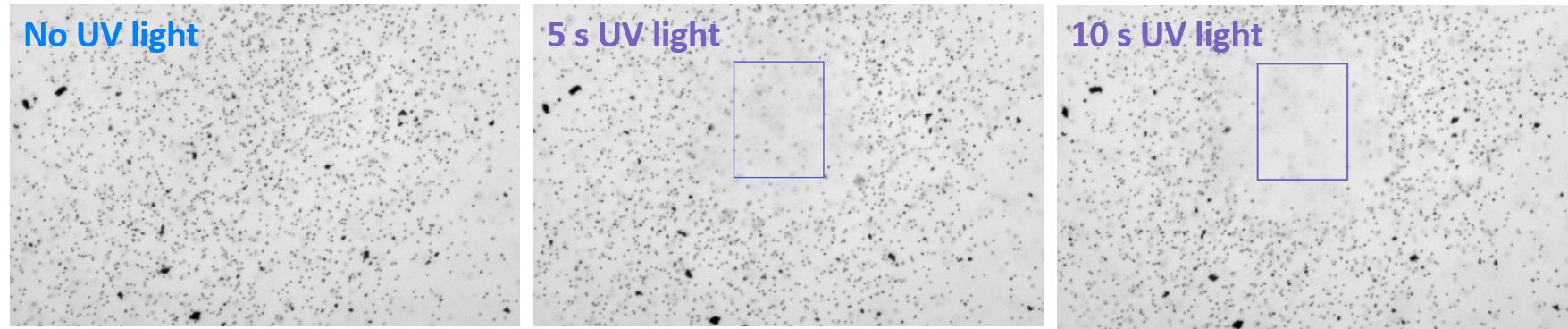}
\caption{\ce{TiO2} particles without surfactant on a solid
  substrate. Left: before UV irradiation. Middle: after 5 s of UV
  irradiation. Right: after 10 s of UV irradiation.}
	\label{nosurfactantsolid}
\end{figure}

\begin{figure}[!thb]
\centering
\includegraphics[width=0.99\columnwidth]{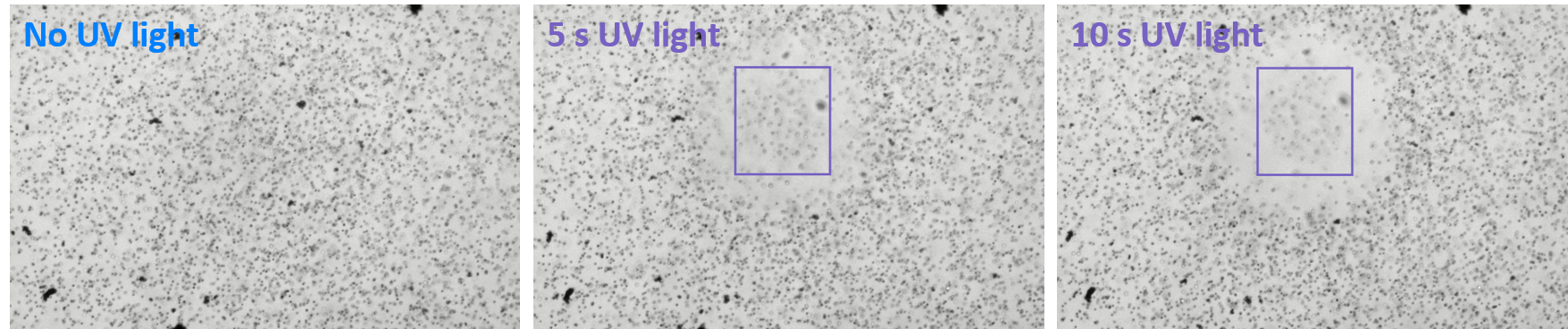}
\caption{\ce{TiO2} particles in a solution of TritonX on a solid
  substrate. Left: before UV irradiation. Middle: after 5 s of UV
  irradiation. Right: after 10 s of UV irradiation.}
\label{surfactantsolid}
\end{figure}

Similarly to the observations in the case of \ce{TiO2} particles on the 
oil--water interface without surfactant addition, when the interface is replaced 
by a wall the irradiation with UV light causes solely an enhanced Brownian 
diffusion and an upwards motion of the particles, irrespective of the absence 
(see Fig.~\ref{nosurfactantsolid}) or presence (see Fig.~\ref{surfactantsolid}) 
of surfactant.

\section{Theoretical model and analysis}
\label{app:theory} 

Here we solve Eqs.~(\ref{eq:cont}) and (\ref{eq:nablau}), under the assumption 
of radial symmetry, in order to compare with the observations. We take the 
initial distribution of particles in the monolayer to be  homogeneous, with 
density $\varrho_0$, and we approximate the illuminated area as a disk of fixed 
radius $L$. Therefore, the evolved density field will have radial symmetry about 
the center of the illuminated area. The velocity field of the Marangoni flow, 
which will only have a radial component $u_r(r,t)$, can be obtained easily by 
applying Gauss theorem to Eq.~(\ref{eq:nablau}): thus, the flow outside of the 
illuminated area is given by
\begin{equation}
  \label{eq:ur}
  u_r(r,t) = \frac{G N(t)}{2\pi r}, \qquad r>L ,
\end{equation}
where $N(t)$ is the number of particles inside the illuminated area. By 
applying Gauss theorem over the illuminated area to Eq.~(\ref{eq:cont}) and 
combining with Eq.~(\ref{eq:ur}), one obtains
\begin{equation}
  \label{eq:dNdt}
  \frac{\partial N}{\partial t} = - 2\pi L u_r(L,t) \varrho(L,t) 
  = - G N \varrho(L,t) .
\end{equation}
Assuming $G<0$, the velocity field~(\ref{eq:ur}) describes a ``falling flow'' 
into the illuminated region, so that all the particles approaching the rim of 
this region, and which determine the value $\varrho(L,t)$ of the density there, 
come from the non-illuminated region. But the 2D flow is incompressible in this 
region (see Eq.~(\ref{eq:nablau})), and consequently, the density at the rim 
does not change in time and it is equal to the initial one\footnote{More  
specifically, in the non-illuminated region the material derivative of the 
monolayer density vanishes,  $d\varrho/dt = \partial\varrho/\partial t + 
\vec{u}\cdot\nabla\varrho=0$, according to Eq.~(\ref{eq:cont}) with 
$\nabla\cdot\vec{u}=0$.}, $\varrho(L,t)=\varrho_0$. Equation~(\ref{eq:dNdt}) 
then renders
\begin{equation}
  \label{eq:N}
  N(t) = N_0 \mathrm{e}^{t/T},
\end{equation}
where $T = 1/(|G|\,\varrho_0)$ is a characteristic time scale and $N_0 = \pi L^2 
\varrho_0$ is the initial number of particles inside the illuminated area. 
Combining Eqs.~(\ref{eq:N}) and (\ref{eq:ur}), one arrives at the prediction in 
Eq.~(\ref{eq:fieldu}) with $A=\pi L^2$.

The distance $R(t)$ of a particle to the center can be obtained by solving the 
equation of motion $dR/dt = u_r (R,t)$ with the initial condition $R(t_i)=R_i$. 
The initial velocity is given by Eq.~(\ref{eq:Vi}), and the correction thereof 
can be estimated by Taylor--expanding the velocity $V(t) = u_r(R(t),t)$ around 
the initial time:
\begin{equation}
  \label{eq:Verror}
  \frac{V-V_i}{V_i} \approx \frac{1}{V_i} \left.\frac{d
      V}{dt}\right|_{t=t_i} (t-t_i)
  = \left( 1 - \frac{V_i T}{R_i} \right) \frac{t-t_i}{T} .
\end{equation}
With the value $T\approx 11\,\mathrm{s}$ obtained from the fit in
Fig.~\ref{HighConc}F, this expression predicts that the relative error
of approximating $V\approx V_i$ is at most $\approx 10\,\%$ for the
trajectories depicted in Fig.~\ref{HighConc}E.

\newpage


\end{document}